\documentclass[%
 reprint,
 amsmath,amssymb,
 aps,
]{revtex4-2}

\usepackage{graphicx}
\usepackage{dcolumn}
\usepackage{bm}

\usepackage[utf8]{inputenc}
\usepackage[normalem]{ulem}
\usepackage{amssymb,amsmath,amsthm}
\usepackage{MnSymbol}
\usepackage{float}
\graphicspath{{./}}

\usepackage{color}
\usepackage{subfigure}
\usepackage{soul}
\usepackage{enumerate}

\begin{document}

\preprint{APS/123-QED}

\title{Planar network statistics for two-dimensional rupturing foams}

\author{Joseph Klobusicky}
 \email{joseph.klobusicky@scranton.edu}
\author{Elif Onat}%
\affiliation{%
Department of Mathematics, The University of Scranton, Scranton, Pennsylvania 18510, USA
}%

\author{Vasilios Konstantinou}
\affiliation{Lam Research, Tualatin, Oregon 97062, USA
}%

\date{\today}

\begin{abstract}
    We conduct experiments on a class of two-dimensional semiwet foams generated through compressing a three-dimensional soap foam between two glass plates.  To induce a spatially uniform rupturing process on foam boundaries, an additional plate is heated and placed on top of the unheated plates.  For 30 separate foam samples, we record network statistics related to cell side numbers and areas as the foam coarsens over a half-minute.  We find that the Aboav law and a quadratic Lewis Law, two commonly used relations between network topology and geometry, hold well for preheated foams.  To track how well these laws are maintained as the foam ages, we introduce metrics for measuring a foam's disorder over time and build simple autonomous models for these metrics.  While the quadratic Lewis Law is found to hold well  throughout the rupture process, the Aboav law breaks down rapidly when the Gini coefficient, used for measuring disparity of cell areas, is approximately 0.8.  
\end{abstract}

\maketitle


\section{Introduction}

A common topic in materials science is the study of microstructure and its evolution under various coarsening methods. For planar network microstructure appearing in polycrystalline metals, porcelains, and foams,  the most widely researched coarsening process occurs through the continuous evolution of grain boundaries \cite{roth2013bubble,magni2013motion,guidolin2023viscoelastic,fausty20202d,kim2017mean, yanagisawa2023cross}.  The coarsening of metals is induced by annealing, and has a direct relation to its tensile strength \cite{petch1953cleavage,hall1951deformation}.  For foams, coarsening is driven by the transfer of gas between cells with unequal pressures \cite{weaire2001physics}.  In two-dimensional planar networks, topological changes are triggered through two event types.  From the von Neumann-Mullins $n-6$ rule, cells with fewer than six sides will shrink at constant rates proportional to their number of sides minus six, eventually shrinking to a point \cite{von1952discussion,mul56}.  When this occurs, the network maintains its trivalent structure through introducing edges in neighboring cells, known as a T2 move.  An individual edge can also shrink to a point and trigger a T1 move which induces topological changes to its four neighbors \cite{weaire1984soap}.  

\begin{figure*}
\centering
\includegraphics[width=.8\textwidth]{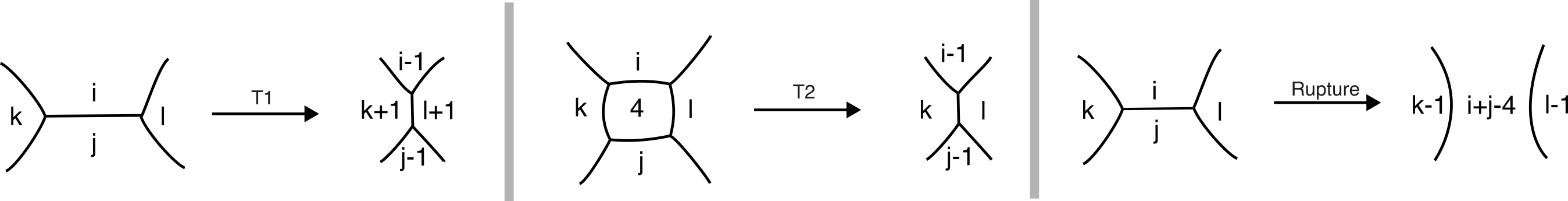}
\caption{Example of a T1 move, T2 move, and edge rupture. The quantities $i,j,k,l$   and $4$ in a cell denote its number of sides. }\label{movesfig}
\end{figure*}

In this study, we study the coarsening process driven exclusively by the rupture of cell boundaries.  This process occurs on a much faster scale than gas diffusion, and is typically encountered in day to day encounters with foams.  Ruptures produce topological reactions which are markedly different from those found in T1 and T2 moves.  Denoting $C_n$ for a cell with $n$ sides, the reaction for the change of topology of the four neighboring cells $(C_i, C_j, C_k, C_l)$ bordering a rupturing edge is (typically) given by the three sub-reactions
\begin{align}
&C_i + C_j  \rightharpoonup C_{i+j-4} &&\textrm{ (Face-merging)}, \label{topreact}\\ 
 &C_k \rightharpoonup  C_{k-1} , \quad C_l \rightharpoonup   C_{l-1} &&\textrm{ (Edge-merging)}.  \nonumber
\end{align}
See Fig.\ \ref{movesfig} for a schematic of possible transitions in a foam.

The merging of the $i$ and $j$ sided cells into a single cell is an example of a second order reaction in which two reactants combine to form a single product, with similarities to  the sticky particle models of Smoluchowski \cite{smoluchowski1916drei}.  Here, clusters $A_i$ and $A_j$ with sizes $i$ and $j$ observe the reaction $A_i + A_j \rightharpoonup A_{i+j}$.  The topological reaction Eq.\ (\ref{topreact}) was the basis for graph and mean-field models of foam rupturing in Ref.\ \cite{klobusicky2021markov}.  In these models, the reaction rate for face-merging resembled the Smoluchowski equation with a multiplicative kernel, which is well-known to produce gelation behavior, or the generation of a massive, infinite-sized cluster \cite{aldous1999deterministic}.  

Several previous experimental studies for foams focused on the evolution of network statistics driven through gas diffusion \cite{duplat2011two,stavans1990temporal,stavans1993evolution,chieco2021experimentally, roth2013bubble}.   The dynamics of rupturing two-dimensional foams, however, is less studied.  An introductory study was undertaken by Burnett et al.~\cite{burnett1995structure} in which bulbs used in a light box served as a natural way to heat the foam.   For this study, the chamber between two plates is filled with a soap solution, sealed, and then vigorously shaken to produce a wet foam.  The liquid is allowed to drain to produce a foam of desired wetness.  In our study, which is detailed in Sec.\ \ref{sec:exp}, a three-dimensional foam sample is placed on top of a plate and compressed with another identical plate.  This produces a foam which is immediately ready for heating, which we refer to as ``semiwet".  We use this term because boundaries are thin enough for bubbles to be approximately polygonal, as opposed to wet foams which have circular cells.  However, the foam is not subjected to draining, which is necessary for producing a dry foam with a low liquid fraction. 

After compressing the foam, another heated plate is placed atop the two plates to produce a rupturing process which typically stabilizes within thirty seconds.   This streamlined process allowed us to obtain multiple samples, a total of 30 experiments with 30 snapshots, taken once per second, producing a dataset of 900 foam snapshots. With these samples we can now produce standard errors and confidence bands for the multiple statistics used to analyze foam properties.  To account for open regions near the foam's border, we impose an artificial circular boundary which serves as a wall for bordering cells.

\begin{figure*}
\centering
\includegraphics[width=.7\textwidth]{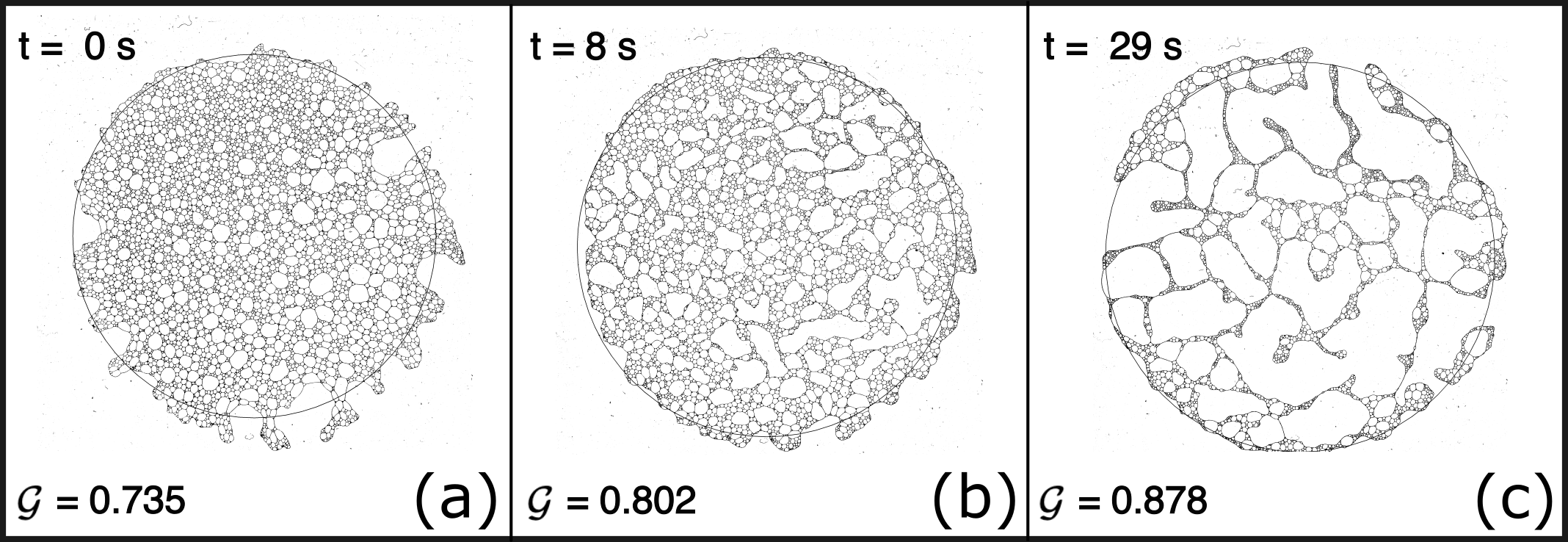}
\caption{Snapshots and Gini coefficients for a foam sample with artificial circular boundary overlaid. (a): Initial conditions before heating with 5512 total cells.  (b): The foam in its intermediate stage after 8 seconds with 4965 total cells.  (c): The aged foam after 29 seconds, with 3222 total cells remaining.} \label{realfoam}
\end{figure*}

In Sec.\ \ref{sec:gen}, we report on general observations produced from the experiments.  The 30 preheated foam samples have variable initial conditions, both in total number of cells and distribution of cell areas.  This is in contrast to studies such as \cite{bae2019controlled}, which is able to design foams with prescribed lattice structures through an intricate system of vacuums.  We also observe that rupturing produces massive, irregularly shaped regions, a phenomenon both observed by Burnett \cite{burnett1995structure} and produced in the computational study of one of the authors \cite{klobusicky2021markov}.  See Fig. \ref{realfoam} for snapshots of a single foam in the rupturing process with circular boundaries overlaid.  In Sec.\  \ref{sec:init}, we conduct a more detailed analysis of foams before heating.  We find uniformity in certain statistics such as the statistical topology of cells.  We also show strong fits with the Aboav linear law \cite{aboav1970arrangement} and a quadratic version of the Lewis law \cite{lewis1926effect}, which are empirical laws used for summarizing the topology and geometry of a planar network.

In Sec.\  \ref{sec:dyn} we model the evolution of a foam's disorder through autonomous differential equations.  The time scale used in these models is internal, using the \textit{rupture fraction}, or the ratio of remaining cell number over initial cell number. A similar approach is also taken in \cite{burnett1995structure}.  For measuring disparity in cell areas, we use the Gini coefficient \cite{gini1936measure}, a quantity typically used in measuring income and other economic disparities.  We also present a measure for topological defect through the second moment of foam's side number distribution minus six.  We find that the Gini coefficient has more desirable properties at measuring network disorder, and it is against this measure that we compare how the Aboav and Lewis laws hold as a foam ages.  

In Ref.\ \cite{klobusicky2021markov}, phase transitions of mean-field models can be readily identified by simply tracking the side number of the largest cell.  For the experiments discussed in this paper, multiple large cells grow in a continuous manner, and it is difficult to distinguish if and in what sense a phase transition occurs. In Sec.\  \ref{sec:fid}, we observe a rapid decay of the correlation coefficient for the Aboav law when the Gini coefficient is approximately 0.8.  This cutoff value can be interpreted as a critical value for a phase transition between ordered and disordered foams. The Lewis law, on the other hand, holds well throughout the rupturing process.  In Sec.\  \ref{sec:dis}, we discuss the implications of these results for modeling foams computationally and measuring the disorder for foams found in manufacturing and industry.

\section{Experimental methods} \label{sec:exp}

%
%
%
%
%
%
%
%
%
%
%

A soap solution is created using 1 teaspoon ($\sim4.93$ ml) of Palmolive brand liquid dish detergent and 500 ml of water in a 1 liter container.  A lid is placed on the container and shaken vigorously to create a three dimensional soap foam which sits atop the liquid solution.  A foam sample of  approximately 30 cubic centimeters is scooped from the container and placed upon a glass plate with dimensions of $20 \times 25 \times 0.24$ cm${}^3$, which is laid on top of a Cricut LED light box.   Spacers of height 0.5 mm are then added to the corners of the plate, and a second plate (with same dimensions as the first) is then slowly placed on top of the first plate.  We note that the height of the spacers are critical to producing foams amenable to analysis.  Spacers which are too tall produce foams with more than a single layer of cells, and those too short cause issues with overly thin Plateau borders which rupture immediately when heated. Adding a second plate compresses the three-dimensional foam to form a single-cell, or quasi two-dimensional, structure.  When compressed, the foam sample spreads to form a structure which is approximately circular.  At locations where the foam meets the upper and lower plates, cell boundaries thicken slightly, but individual cells can still be clearly distinguished when viewed from above.  The foam sample in this state has a radius of about 15 cm, and there are typically around 5000 total cells in the network before any heating is applied. 
    
The foam at this point has a liquid fraction of approximately 15\%.  It is also quite stable, with an occasional rupture occurring every few seconds.  No heat is emitted from the LED light box, so we instead choose to heat from above by using a heated third plate.  This plate is heated uniformly with a heating pad to $45^\circ$C, and is then slowly placed on top of the two plates enclosing the foam.  Recall that spacers are placed between the plates, so adding the third plate does not further compress the foam.  The chamber is not sealed from the sides, but samples are small enough so that no foam escapes from the plates after compression. The foam, however, does change in size, at first shrinking when heated, and then slowly expanding as the plate cools.  This is due to the changing thickness of cell walls which can be explained through surface tension.  For an isotropic soap foam, surface tension is proportional to the total perimeter of cell boundaries.  As temperature increases, surface tension decreases by thickening cell boundaries and subsequently reducing total perimeter, resulting in a uniform shrinking of the foam. 

 During the heating of the foam, cells shrink but few ruptures occur.  Once the foam reaches its smallest size, cells walls become weak enough from the heating for the rupture process to begin.  The process begins slowly, and then quickly speeds up to several hundred ruptures per second. For all foams, after 30 seconds has elapsed, the occurrence of ruptures again becomes infrequent.  Snapshots of the process are taken with a Sony Alpha 7 II camera with pixel resolution $4000 \times 6000$ pixels for an image of size $16.8 \times 25.2$ cm${}^2$.  The camera is placed above the foam at a distance of 80 cm. Using an intervalometer, a snapshot is taken at $t = 0, \dots, 29$ seconds to give 30 snapshots for the rupturing process, with the first snapshot taken after the initial shrinking of the foam.  This experiment is repeated for 30 different foam samples, for a total of 900 foam snapshots.

\subsection{Image processing} \label{sec:img}

In the analysis of cell areas in Ref.\ \cite{burnett1995structure}, large empty spaces between cells are not considered. For the metrics used in Section \ref{sec:dyn},  large gaps between cells are definitive in measuring a foam's disorder, so we will consider all regions as individual cells. With a free boundary, however, an issue arises when trying to quantify the size of large, sometimes labyrinthine, regions created from ruptures, but which are technically part of the foam's exterior.  Our approach for including these regions is to create an artificial circular boundary.  Choosing a circle as a boundary shape is a natural choice, as the foam maintains a roughly circular shape throughout the rupturing process.  The algorithm of creating the foam boundary is as follows:

\begin{enumerate}[i]
\item Compute the centroid $(\bar x, \bar y)$ of cell boundary pixels in the image.
\item Determine a minimal radius $R$ such that at least 90\% of all pixels are contained in the disk centered at $(\bar x, \bar y)$.  
\item Crop the image by inserting an artificial circular boundary centered at $(\bar x, \bar y)$ with radius $R$ and removing all pixels outside of this circle.  \end{enumerate}
 The regions sharing a border with the boundary are now considered as distinct cells.  The circular arcs on the  boundary are counted as cell edges.  We have selected to include 90\% of cells since  the intersection between the outer boundary of the  foam and the circle is minimal, avoiding the creation of abnormally large cells which wrap around the boundary.

Several image processing and morphology packages from the Open CV Library \cite{itseez2015opencv} were employed in Python to process image data from the cropped images.  Adaptive thresholding is applied to the image to account for any imbalances in lighting.  The image is then binarized, followed by an opening (erosion followed by dilation) operation to remove spurious pixels, and then dilated again to thicken the boundaries between cells.  This prevents the identification of two cells as a single connected component.  Dilation also prevents the counting of Plateau border regions found at the foam's triple junctions.  For the semiwet foams we consider, the Plateau borders sometimes contain small, Apollonian-like cells, which are also coarsened out from dilation.  We mention that studies have begun to consider these ``inner cells" in wet foams \cite{galvani2023hierarchical,sauerbrei2006apollonian,kwok2020apollonian}.  Dilating boundaries increases edge thickness, and subsequently decreases cell areas, but since the perimeter of the entire foam is uniformly enlarged, this operation  has no discernible effect when comparing relative areas of cells.

Edge detection algorithms then determine connected components of the binarized image, from which areas can be found readily by a simple pixel count (with a conversion factor of 1 pixel = $42\times 42 \hbox{ } \mu$m${}^2$).  A typical cell boundary has a thickness of about 5 pixels, or $200 \hbox{ } \mu m$, and on average there are approximately 400 pixels per total cell boundary. For each cell, the number of neighbors (or sides) were found by determining connected components (faces) of the planar network, and constructing an adjacency matrix between the cells.  This enables us to determine first-order topological correlations used in finding fits for the Aboav law.

\section{Results}

\subsection{General observations and an internal timescale}  \label{sec:gen}

In Fig.\ \ref{realfoam}, we show a sample foam at three stages in its evolution, with and without the artificially imposed circular boundary.  In its initial conditions, the foam is somewhat uniform in its area and side distributions, although we still observe several large cells with over fifty neighbors.  While gelation behavior, which we define for this study as the formation of massive cells with many sides, occurs at different times for each foam, we note for this sample foam that by the middle of the experiment, at 8 seconds, a multitude of large cells have formed. At this point, the Gini coefficient, a measure for coarsening that we examine in Sec.\ \ref{sec:gen}, is approximately 0.8.  As shown in Sec.\ \ref{sec:fid}, at this value the Aboav linear law begins to break down. Merging of both large and small cells continues until the end of the experiment at 29 seconds, where the interiors of the 20 largest cells comprise most of the enclosed disk.  Over all samples, the most evident observation for rupturing foams is the rapid generation of multiple large regions.  As mentioned in Sec.\ \ref{sec:img}, we consider these regions as proper cells. As time progresses, these regions grow and are, in general, irregularly shaped and nonconvex.  Surrounding these large regions are smaller, convex cells, with thickening Plateau boundaries as the liquid content of the foam is distributed to a smaller amount of cells.  These smaller cells often form thin, bridgelike structures between the massive cells.

The rupture rate for foams varies across samples.  Some foams begin rupturing almost immediately after their shrinking period, and generate massive cells within five seconds.  Others take several seconds before beginning a slow rupturing process producing multiple large cells.  The timing for ruptures, in general, has been found to be erratic \cite{vandewalle2001cascades}, generally occurring in cascades with nontrivial spatial correlations.   This is in contrast to coarsening under gas diffusion, where numerical and physical experiments demonstrate the annihilation of cells at a linear rate which occurs  uniformly across the foam \cite{elsey2011large}.  A spatial correlation of ruptures in our experiments also appears to exist, although for foams with a large number of ruptures, massive cells are approximately uniformly distributed across the circular domain by the end of the rupturing process.

\begin{figure}
\centering
\includegraphics[width=.5\textwidth]{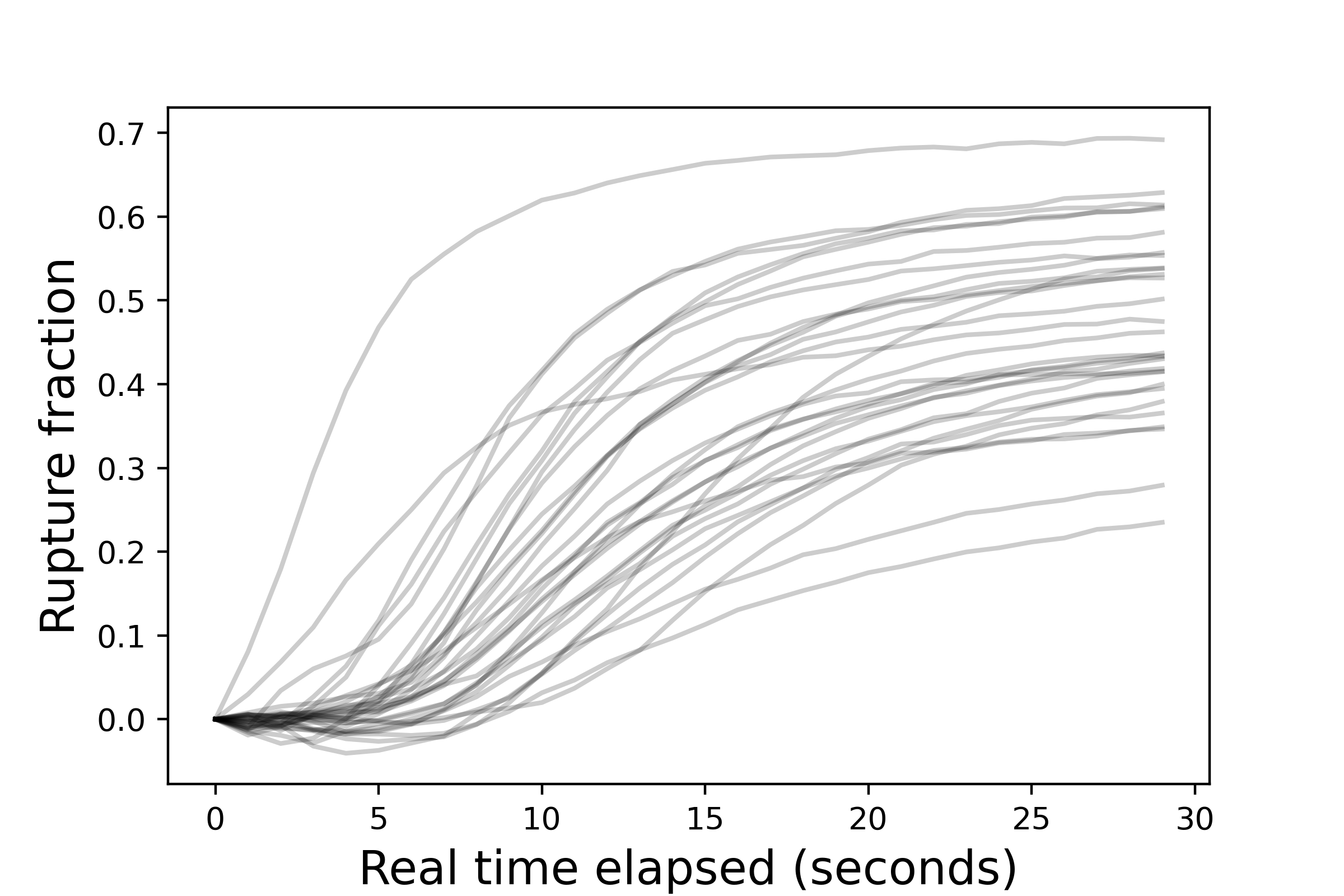}
\caption{The rupture fraction for 30 foam samples, with snapshots taken at $t = 0, \dots, 29$ seconds. }   \label{realtimefoam}
\end{figure}

To use a time scale which is more amenable to dynamic modeling, we will work with an ``internal clock" of the foam, in which we track total cell numbers relative to  initial conditions.  In nearly all ruptures, a single rupture follows the reaction Eq.\ (\ref{topreact}), and decreases the total cell count  by one. We will call this time scale the \textit{rupture fraction} as the total number of ruptures is approximately the total reduction in cells (see Ref.\ \cite{klobusicky2021markov} for some counterexamples where ruptures can remove more than one cell). For a planar network $G_t$ and $ |G_t|$ denoting the total number of cells after $t$ seconds, we define the rupture fraction as
\begin{equation}
s(t) = 1 - \frac{|G_t|}{|G_0|}. \label{rupfrac}
\end{equation}
In Fig.\ \ref{realtimefoam}, we plot $s(t)$ for $t = 1, \dots, 29$ for each of the 30 foam samples. Note that in some slower rupturing foams, we observe a ``time reversal" near the beginning where movement of smaller cells near the artificial circular boundary causes a slight increase in the total number of cells in the circular region.  For the purposes of this study, in which we are more interested in the behavior for highly aged foams, this effect is minor in terms of modeling considerations.

\subsection{Statistics of initial conditions}\label{sec:init}

\begin{figure}
\centering
\includegraphics[width=.5\textwidth]{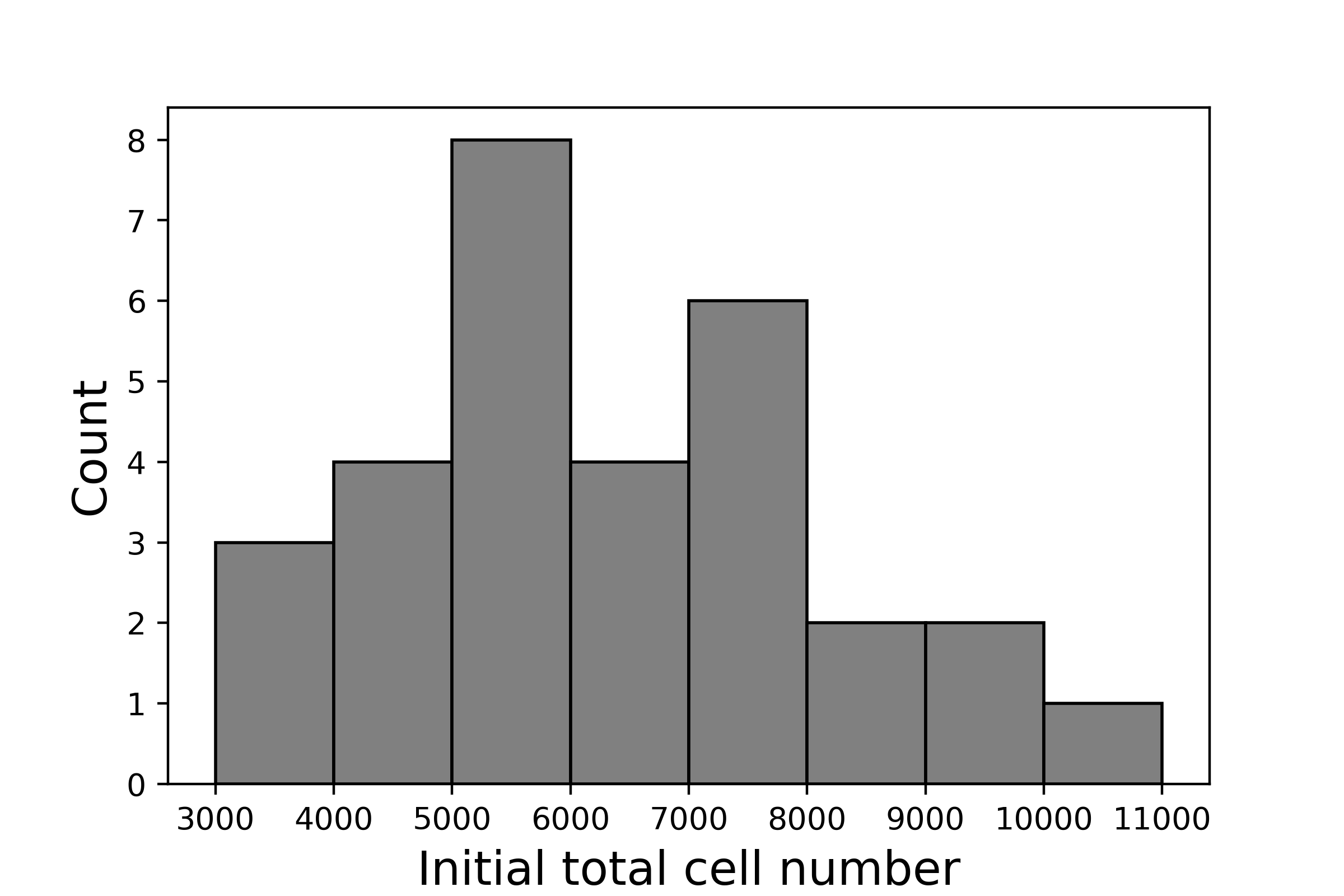}
\caption{A histogram of total cells taken before the heating process for 30 foam samples.}   \label{foamhist}
\end{figure}

The three-dimensional foams created from shaking a foam solution vary in wetness and average cell size.  Foam samples are taken from near the liquid/foam interface, where cell boundaries are thicker and bubble volumes are smaller.  The number of initial cells ranges from 3381 to 10743. A histogram of total initial cell numbers is given in Fig.\  \ref{foamhist}, and the variation of wetness produces differing rates of rupture as shown in Fig.\ \ref{realtimefoam}.  While we find variation among the samples in terms for total initial cell numbers and wetness, we show in this section we will find that these foams have several similar network statistics.

\subsubsection{Side number and area distributions}

\begin{figure*}
  \centering
  \subfigure{\includegraphics[width=.45\textwidth]{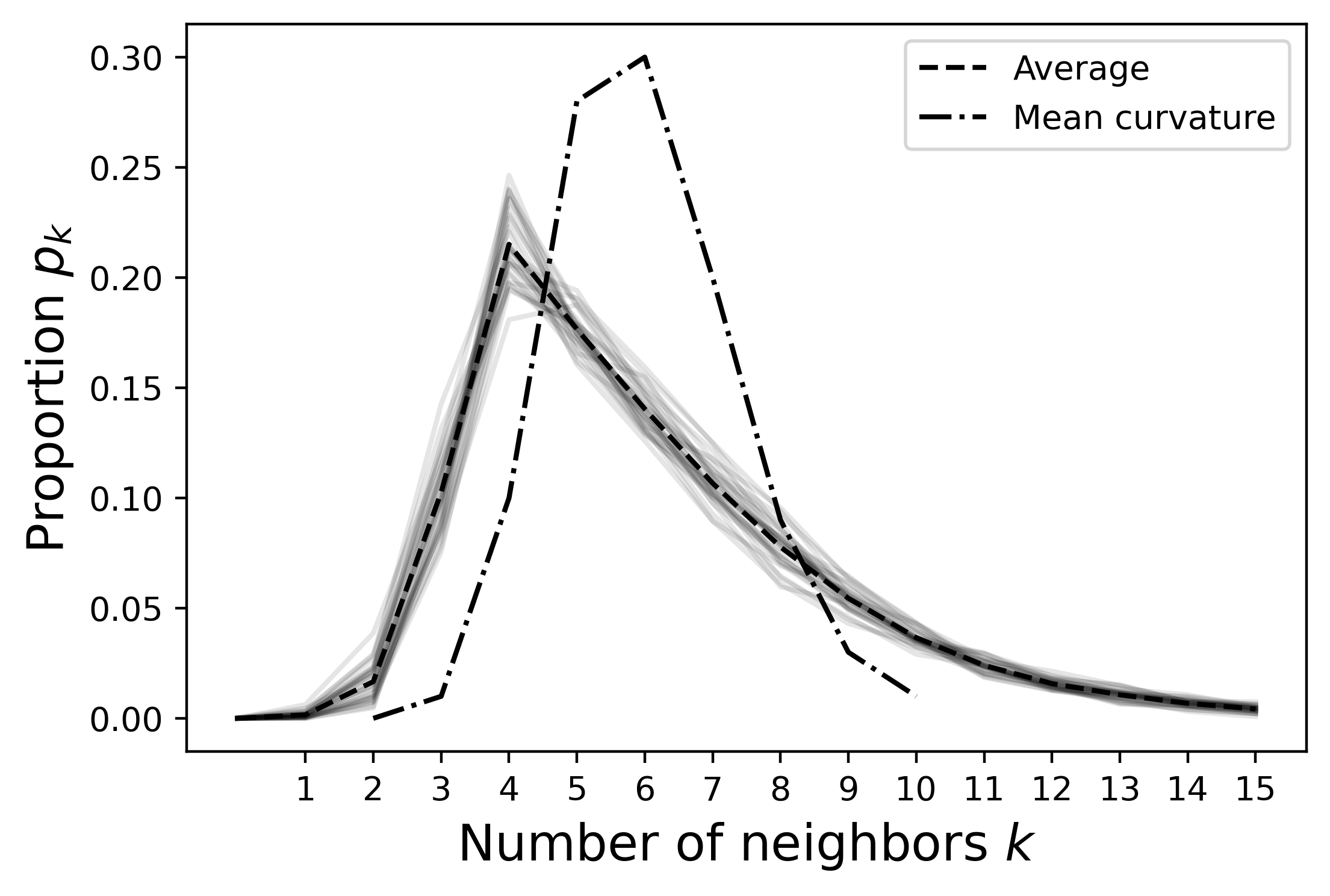}}
  \subfigure{\includegraphics[width=.45\textwidth]{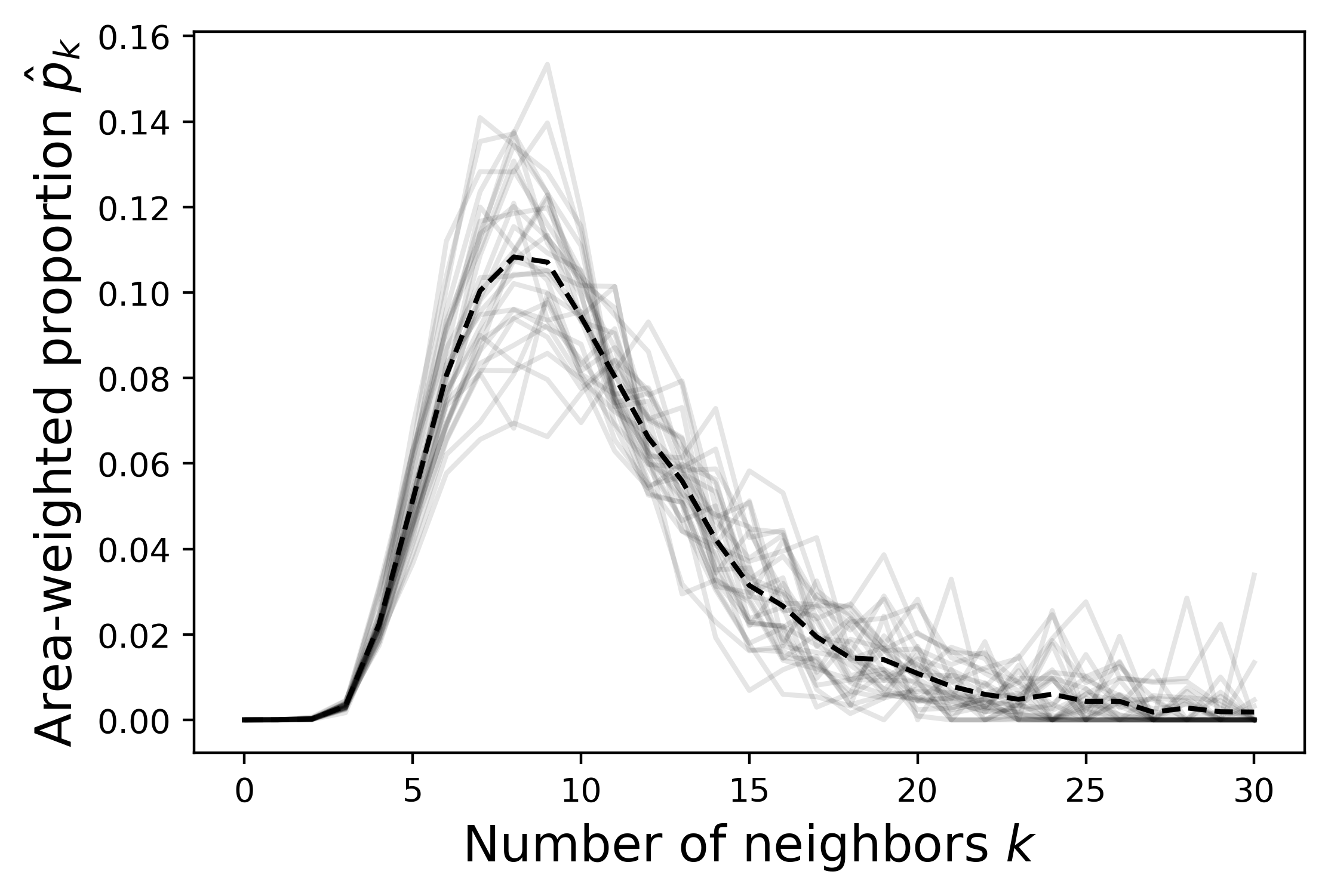}}
  \caption{Side distributions of foams before heating. In both figures, transparent lines correspond to distributions of individual samples, and the averages of these distributions are plotted with a  bold dashed line.  Connecting lines between integer values of neighbors serve as a visual aid.  Left:  Proportion of cells with $k$-neighbors for $k = 1, \dots, 15$.  Also shown for comparison is the stationary distribution for coarsening through mean curvature found in \cite{elsey2011large}.  Right: Area-weighted proportion of cells with $k$-neighbors for $k = 1, \dots, 30$. }

  \label{stattopfig}
\end{figure*}

The side number distribution, also called the statistical topology, gives proportions $p_k$ of $k$ sided cells for $k \ge 1$.   A key observation in coarsening by diffusion is that under a wide range of ordered and disordered initial conditions, the statistical topology distribution converges to a universal attractor, even though the network continues to coarsen with average cell area increasing at a constant rate \cite{elsey2011large}.  In Fig.\ \ref{stattopfig}, we plot the side distribution for the 30 foam samples before the rupture process begins.  As expected with a cubic planar graph (having all vertices of degree three),  the mean number of neighbors $\langle n \rangle =  \sum_{n\ge 1} p_n n$ is approximately 6 , with $\langle n \rangle = 5.9 \pm 0.1$  (we report this and all future confidence intervals with plus or minus one standard error).  Despite the range of total numbers of foams, the side distribution for all of the foams are similar, with the mode of almost all foams occurring at 4 sides.  We contrast this distribution to the universal attractor distribution from numerical studies of coarsening driven by mean curvature flow in Ref.\ \cite{elsey2011large}.  This distribution (also plotted in Fig. \ref{stattopfig}) is more concentrated near its mode of 6 sides, and rarely has cells with more than 10 sides.

We also plot the area-weighted side number distribution in Fig.\  \ref{stattopfig},  where $\hat p_k$ for $k \ge 1$ gives the probability that a randomly selected interior point in a foam is inside of a $k$-gon.  This weighted distribution is used in finding the average coarsening rate from gas diffusion in \cite{roth2013bubble}. As cells with more sides tend to be larger (an immediate consequence of the Lewis law), we should expect, and indeed find, that using area-weighted distributions increase both the mode and tail probabilities.  Denoting the mean area-weighted number of sides as  $\llangle n \rrangle  = \sum_{n\ge 1} \hat p_n n$, we find $\llangle n \rrangle = 10.2 \pm 0.9$.  Distributions across different samples are also more erratic than unweighted distributions.  In particular,  ``spikes" with proportions of several percent frequently arise from a few large cells all having the same side number.

\begin{figure}
\centering
\includegraphics[width=.5\textwidth]{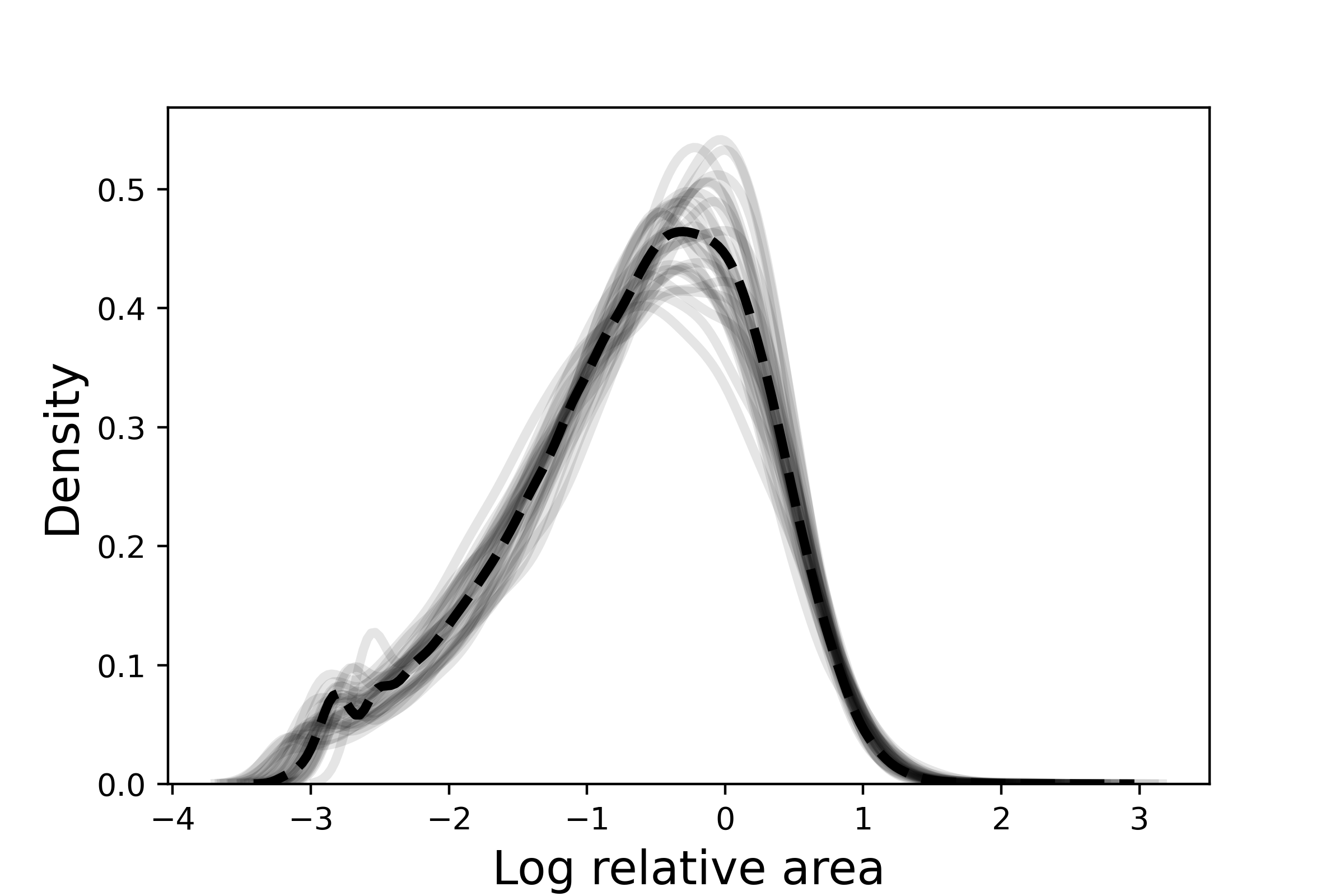}
\caption{Density of log relative areas of foams before heating.  Individual densities are plotted with transparency, and the mean density over all foams are shown with a dashed line. } \label{initsareas}
\end{figure}

We also plot the distribution of cell areas in Fig.\ \ref{initsareas}.  To better visualize the relation between cell sizes, we scale cells areas to have a mean of 1, and plot the base ten log of these areas.  Even before heating, we find that cell areas exhibit  a multiscale behavior, with large cells with many sides surrounded by smaller 3- and 4-gons with areas differing by several orders of magnitude.  Cell areas appear to vary more  than statistical topologies when considered across different foam samples. However, we find that most distributions have roughly the same shape\textemdash generally left skewed, with a mode occurring near the mean cell area.

\subsubsection{Aboav and Lewis Laws} \label{sec:abo}

For a cell with $n\ge 2$ sides, Aboav's law \cite{aboav1970arrangement}  is an empirical observation  that the average number of sides $m_n$ for neighboring cells can be approximated by
\begin{equation}
m_n = a + \frac bn \quad \Rightarrow \quad M_n := n m_n = a n +b.
\end{equation}

Typically,  $M_n$  is plotted instead of $m_n$, and the model's fit is then measured with linear regression.  Computing $M_n$ is a straightforward exercise when equipped with the adjacency matrix from the dual graph of the foam. 

With all 30 samples and considering cells with $n = 2, \dots, 12$ sides, we compute linear regression parameters for $M_n$.  We find a strong correlation, with the average coefficient of determination  $r^2 = 0.98 \pm 0.01$.  The coefficients of Aboav's law have values $a = 5.4 \pm 0.5$ and $b = 23 \pm 4$. 

\begin{figure}
\centering
\includegraphics[width=.5\textwidth]{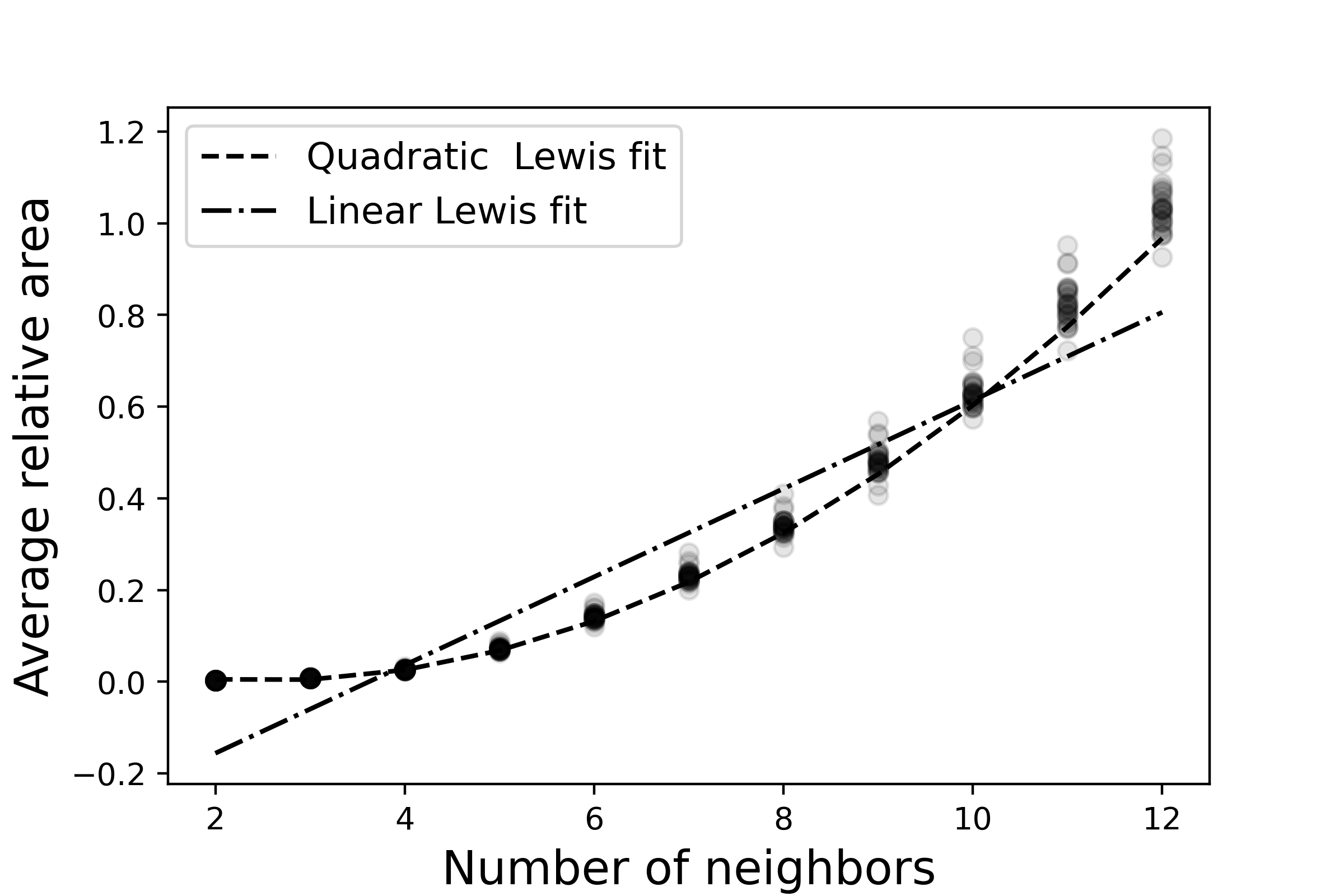}
\caption{Average relative areas $A_n$ of $n$-sided cells for $n = 2,\dots, 12$ over all foam samples plotted with transparency.  The linear and quadratic Lewis law fits use average values of model parameters across all 30 samples.} \label{lewisfig}
\end{figure}

Another empirical law related to cellular microstructure is the Lewis law \cite{lewis1926effect}, which proposes  a linear relationship between side number $n$ and the average area $A_n$ of $n$-sided cells.  For a linear fit $A_n = cn+d$ of relative areas, we find an average coefficient of determination $r^2 = 0.909 \pm  0.003$, but plots for $A_n$ appear to be convex, with a residual analysis suggesting that linear regression is not an appropriate model.  Instead, we consider a quadratic fit $A_n = \beta_2 n^2+ \beta_1 n+ \beta_0$.  See Fig.\ \ref{lewisfig} for $A_n$ plotted across all samples and the average linear and quadratic Lewis laws.  The coefficient of determination for the quadratic model for each foam sample is $R^2 > 0.999$ with coefficient values  $\beta_0 = 0.0116 \pm 0.0005$, $\beta_1 = -0.060 \pm 0.003$, and $\beta_2 = 0.080 \pm 0.007$.

\subsection{Dynamic statistics of gelation}  \label{sec:dyn}

As shown in Fig. \ref{realfoam}, we find that the second order reaction  (\ref{topreact}) induced by edge ruptures creates increasing disparities between massive cells bordering hundreds of cells and small, mostly unaffected three- and four-sided cells. The reassignments of side number are related to those associated with the Smoluchowski equation for sticky particle clusters with a multiplicative reaction kernel.  For this model, at a positive finite time a single cluster experiences an exposive growth producing a massive, infinite-sized cluster called the \text{gel}.  Borrowing terminology for our study, we will refer to the creation of massive cells with many sides as \textit{gelation}.  In creating kinetic models for both cluster models \cite{lushnikov1978some,marcus1968stochastic} and those for coarsening behavior of foams \cite{marder1987soap,flyvbjerg1993model, fra881,fra882,klobusicky2020two, klobusicky2021markov}, all-to-all connectivity is typically assumed.  This assumption leads to gelation behavior which is concentrated in a single cluster or cell.  However, Fig. \ref{realfoam} reveals that gelation in rupturing foams is often shared between several cells, necessitating metrics which gives similar results for when a large area is shared by either one or a small number of massive cells. We develop metrics in this section for measuring gelation from the perspective of area and topology.

\subsubsection{Gelation of area} \label{ginisec}

\begin{figure*}
  \centering
  \subfigure{
    \includegraphics[width=.45\textwidth]{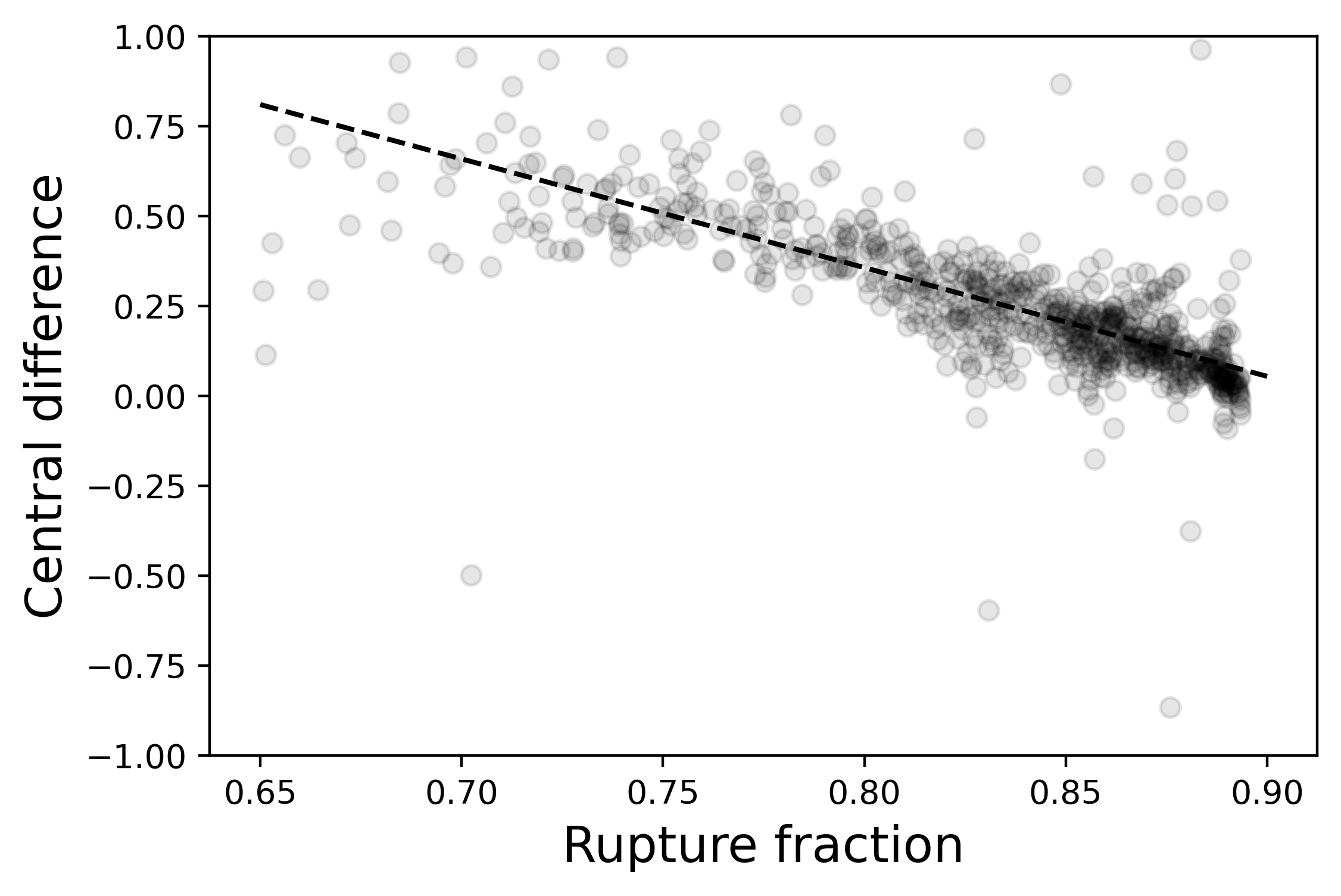}}
  \subfigure{
    \includegraphics[width=.44\textwidth]{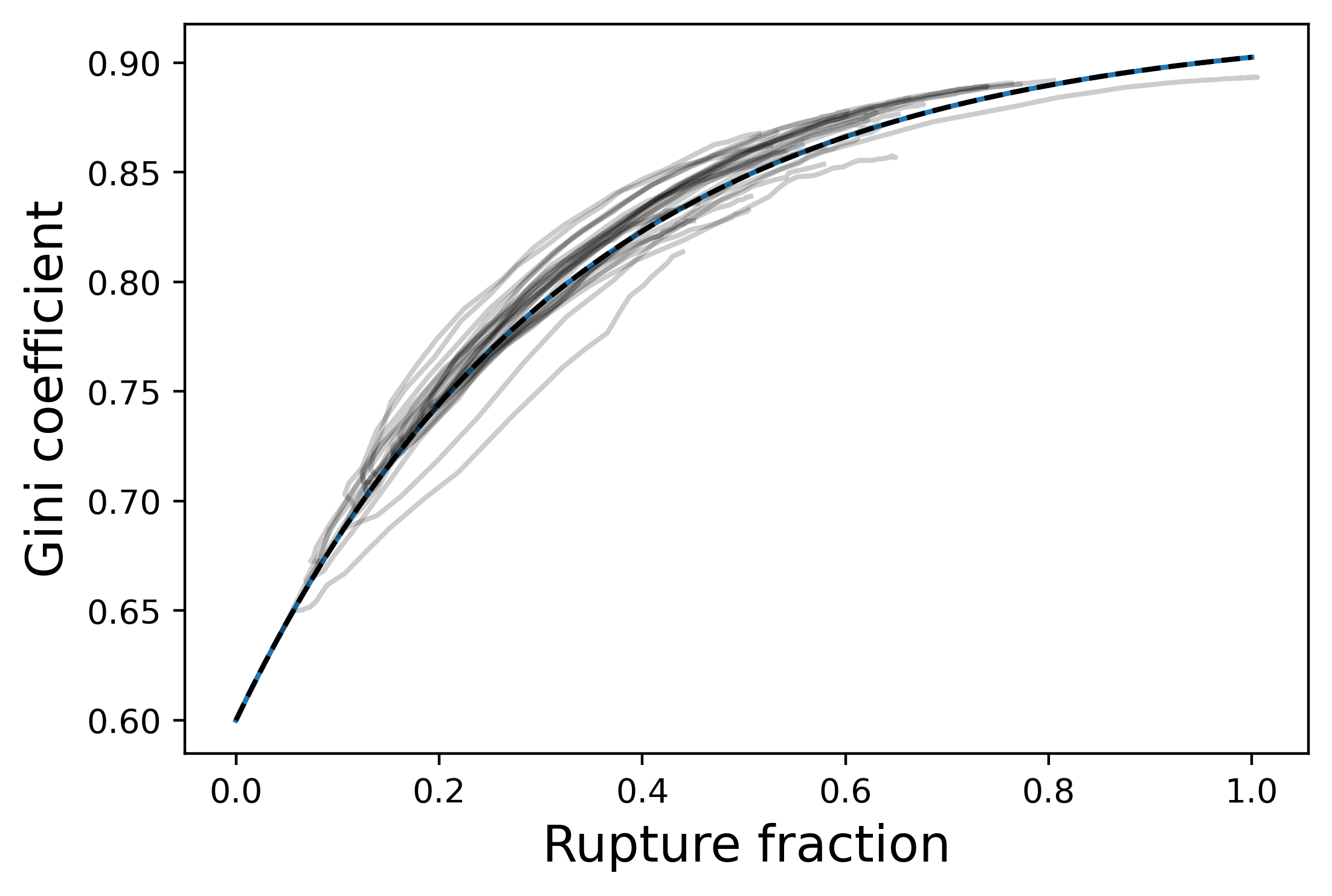}}
  \caption{Curve fitting for $\hat{\mathcal G}(s)$. Left: Linear regression of $\hat \Phi(g) = ag +b$, with $a =-3.02$ and $b= 2.77 $.  Right: The solution curve $\hat{\mathcal G}(s)$ with initial condition $ g_0 = 0.6$.  For each foam sample we also overlay, with transparency, the translated gelation curves $\mathcal{G}_i(s-\hat{\mathcal G}^{-1}(\mathcal G_i(0)))$ for $i = 1 \dots, 30$.}

  \label{stogini}
\end{figure*}

An effective measure for measuring disparities in cell areas is given by the Gini coefficient, originally formulated to compute the income and other economic disparities among populations \cite{gini1936measure}.  We will use this measure to measure disparities among cell areas in a foam.  Specifically, let $A = (a_1, \dots, a_N)$ be a nondecreasing sorted list of cell areas.  For $q \in [0,1]$,  define $g(q)$ by letting $g(k/N) = \sum_{i = 1}^k a_i/\sum_{i = 1}^N a_i$ for $k = 1, \dots, N$ and defining other points in $[0,1]$ through linear interpolation.  The Gini coefficient for cell areas is then 
\begin{equation}
\mathcal{G} = 1 - 2\int _0^1 g(q)dq.
\end{equation}
For a hexagonal lattice on a fixed circular boundary, $\mathcal G$ approaches zero as the number of cells becomes large.  At the other extreme, a single massive cell encompassing almost the entire domain border by many small cells can produce a Gini coefficient arbitrarily close to 1.

Under the assumption that the growth of the Gini coefficient $\mathcal G$ only depends on its value, we can model $\mathcal G(s)$ as an autonomous differential equation, written as 
\begin{equation}
\frac{d}{ds}\mathcal{G}(s)  = \Phi[\mathcal{G}(s)  ], \quad \mathcal G(0) =  g_0,\label{ode}
\end{equation}
where the rupture fraction $s \in [0,1]$ in Eq.\ (\ref{rupfrac}) is used as a time variable. In most cases, we find that the Gini coefficient increases in time, so in general  we should require that $\Phi(g) $ is a continuous, positive function for $g \in [0, g_*)$ with $g_* \le 1$.  

We will now examine one approach for modeling $\Phi$ through regression, using a dataset of central differences to estimate derivatives of $\mathcal G(s)$.  In particular, we compute $(s_i^j, \mathcal G_i^j)$ denoting, respectively, the rupture fraction and Gini coefficient for the $i$th foam taken after $j$ seconds, where $i = 1, \dots, 30$ and $j = 1, \dots, 29$. Derivatives at rupture fractions $\hat s_i^j = (s_i^{j+1} + s_i^j)/2$ are then approximated by
\begin{equation}
	\mathcal G'(\hat s_i^j) \approx  \frac{\mathcal G_i^{j+1} - \mathcal G_i^j}{s_i^{j+1} - s_i^j} . \label{cd}
\end{equation}
Figure  \ref{stogini} shows that the central differences can be reasonably approximated with a linear function $\Phi(g) = ag +b$.  Under such a linear fit, this produces, for initial conditions $\mathcal G(0) = g_0$, the solution curve 
\begin{equation}
\mathcal G(s) = e^{as}\left(g_0+\frac ba\right) - \frac ba.\label{ginisoln}
\end{equation}
Using a linear regression, we find a fit of $a = -3.02$ and  $b= 2.77$. From (\ref{ginisoln}), we use the regression parameters of $\hat \Phi$ with $g_0 = 0.6$ (chosen to be smaller than the Gini coefficient for all foams)  to obtain the fit $\hat{\mathcal{G}}(s)$.

Under the time scale of rupture fraction, each of the foams we have observed start at different initial times.  However, we can use the autonomous property of our model (\ref{ode}) to align times so that the initial conditions of Gini cofficients   all lie on the same solution curve $\hat{\mathcal G}(s)$ with predetermined initial condition $g_0$.  Specifically, given the initial condition $\mathcal G_i(0) = g_0^i>g_0$, we predict
\begin{equation}
	\hat{\mathcal{G}_i}(s) = \hat{\mathcal G}(s+\hat{\mathcal G}^{-1}(g_0^i)).
\end{equation}
 To visualize how the solution curve $\hat{\mathcal G}(s)$ compares against each of the empirical graphs of Gini coefficient $\mathcal G_i(s)$, we plot $\hat{\mathcal G}(s)$  against the translated curves $\mathcal{G}_i(s-\hat{\mathcal G}^{-1}(g_0^i))$  so that initial conditions for empirical curves lie on the solution curve.  We find a reasonable fit between the empirical curves and the model, although for two foams with smaller initial Gini coefficients we observe a considerable lag in growth compared to the model.

\subsubsection{Gelation of topology}

We can also measure gelation of a foam from a topological perspective by computing a distance to an ordered foam containing only hexagons.  To define this measure, we consider the random variable $S$ of the side number of a random cell selected with uniform probability. For the probability mass function $\{p_s\}_{s \ge 1}$ denoting probabilities of selecting $s$-sided cells, we define the \textit{topological gelation} which measures the second moment of the topological defect $S-6$, given by

\begin{equation}
\mathcal T = \mathbb E[(S-6)^2] = \sum_{s \ge 1} p_s (s-6)^2.
\end{equation}
Like the Gini coefficient, $\mathcal T$ is nearly 0 for a perfect hexagonal lattice on a circular boundary. However, unlike the Gini coefficient, topological gelation can be arbitrarily large.  In particular, a foam consisting of a single interior cell of $n$ sides adjacent to a band of $n$ 4-sided boundary cells gives a topological defect that is asymptotic to $n$ as $n \rightarrow \infty$.

\begin{figure}
\centering
\includegraphics[width=.5\textwidth]{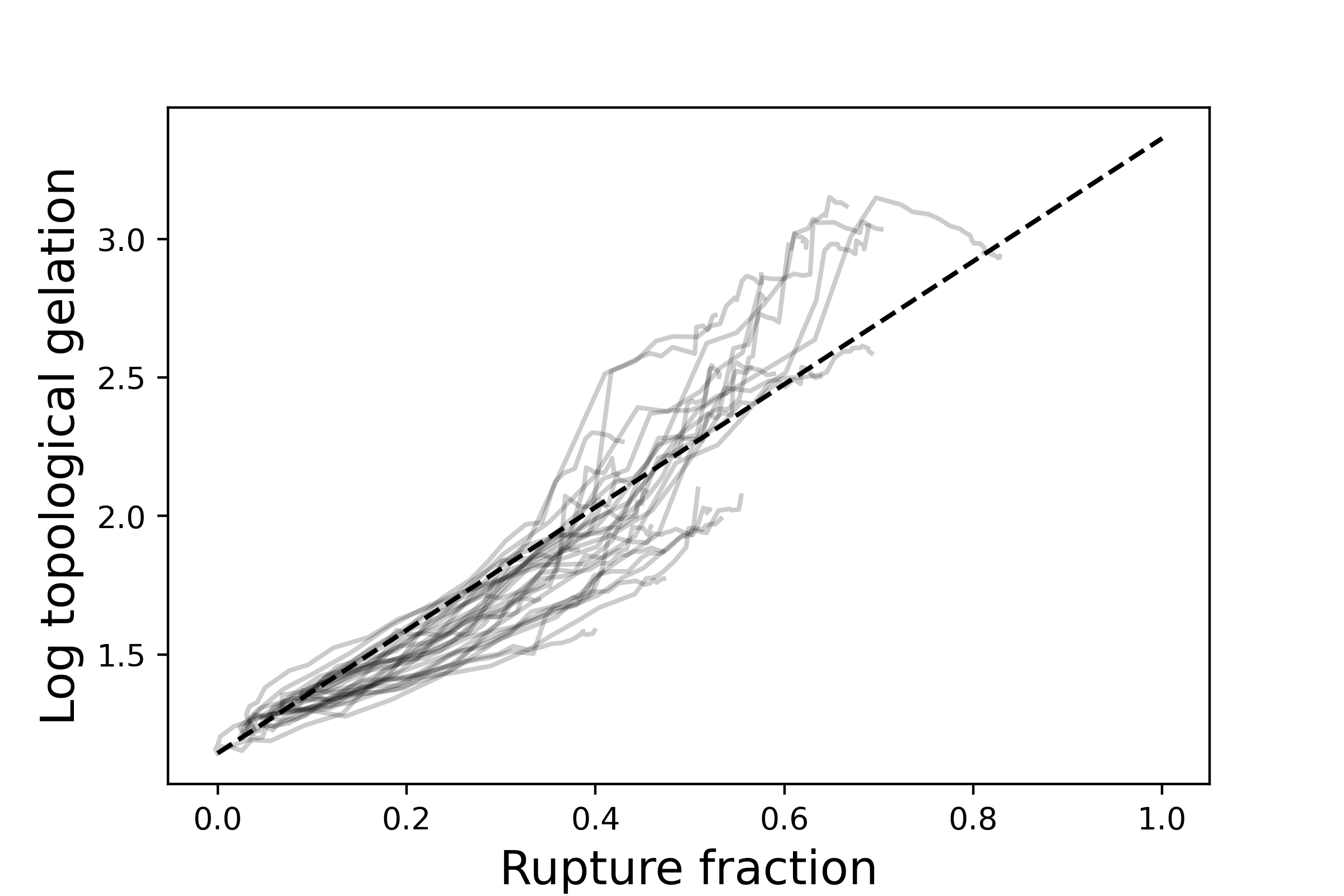}
\caption{The linear fit $\widehat{\ell}(s)  = as+b$ with $a = 2.22 $ and $b = 1.14$ for the log of topological gelation. For each foam sample we also overlay, with transparency, the translated gelation curves $\ell_i(s-\widehat{\ell}^{-1}(\ell_i(0)))$ for $i = 1 \dots, 30$.} \label{logtopfig}
\end{figure}

Topological gelation appears to grow faster than a linear rate, and so to check for power law growth we compute linear fits for log data $\widehat{\ell}_i:= \log_{10}( \widehat{\mathcal T}_i(s)) = a s +b$ for each observed gelation curve $\mathcal T_i(s)$.  The average of these coefficients is computed to give a fit $\widehat{\ell}(s)) = as+b$ with $a = 2.22 $ and $b = 1.14$.  We plot this regression line in Fig. \ref{logtopfig}.  Similar to shifting gelation curves for the Gini coefficient, we assume the model is autonomous and plot $\ell_i(s-\widehat{\ell}^{-1}(\ell_i(0)))$ against the model $\widehat{\ell}(s)$.  It appears, however, that the growth rate in topological defect increases more highly aged foams.  For foams with $\mathcal T<100$, growth scales at approximately $O(s^2)$, while for $\mathcal T>100$, we find a growth closer to $O(s^4)$.  We also note that a few highly aged foams sometimes decrease in $\mathcal T$.  This can be an effect of several artifacts, including the movement of foam pockets into the artificial boundary, and the fusing of two regions of bubbles to reduce total variance of cell neighbors.

\subsubsection{Decay of fit for Aboav and Lewis laws}  \label{sec:fid}

\begin{figure*}
\centering
\includegraphics[width=\textwidth]{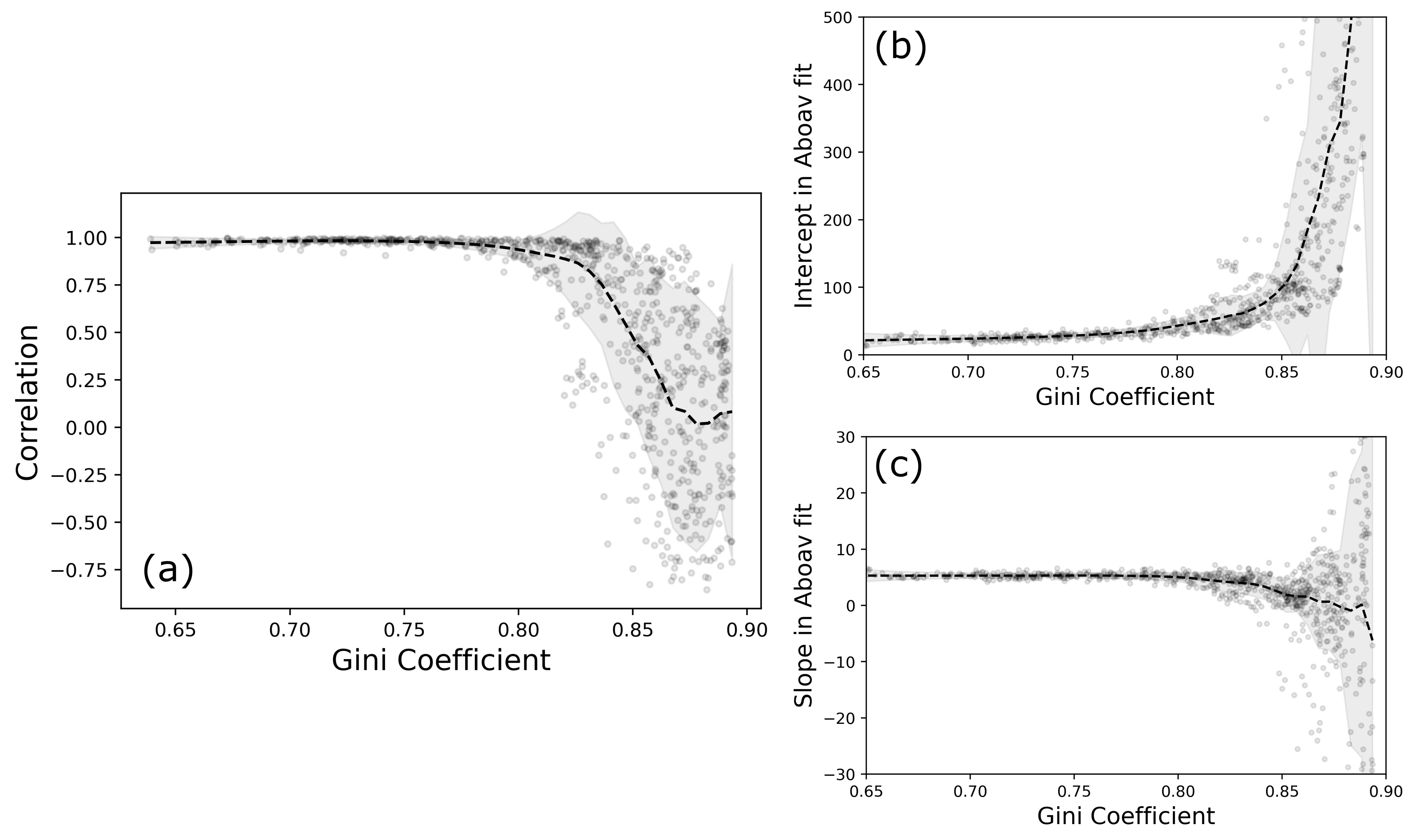}
  \caption{Scatterplots and LOWESS regression curves for Aboav's Law $M_n = an+ b$ over all foam snapshots, comparing the Gini coefficient $\mathcal G$ with  (a) correlation coefficient, (b) value for intercept parameter $b$, and (c) value for slope parameter $a$. For each image, the 95\% bootstrap confidence band is shown in light grey. }
  \label{aboavevo}
\end{figure*}

In Sec.\ \ref{sec:abo}, for unheated foams we found a strong coefficient of determination value for both the linear Aboav law and quadratic Lewis law.  As the foam ages, it is not clear if the parameters in these models remain constant, or even if these models keep their high values of $R^2$.  From Sec.\ \ref{sec:dyn}, we find that the growth of the Gini coefficient is smoother and more predictable than topological gelation, so this will be the metric we will compare against the Aboav Law. In  Fig.\ \ref{aboavevo}, we plot $\mathcal G$ against the correlation coefficient over all snapshots.  We also plot $\mathcal G$ against the fitted parameter values $a$ and $b$ in the Aboav fit $M_n = an+b$.  To create smoothed estimates we use locally-weighted scatterplot smoothing (LOWESS) regression \cite{cleveland1979robust}.  To create a 95\% confidence band, we generate 100 bootstrap samples of LOWESS curves, each derived from randomly selecting half of the data.

We find that the linear Aboav fit holds quite well when  $\mathcal G < 0.8$, with $r = 0.975 \pm 0.001$.  Fitted parameter values are also stable in the regime, with $a = 5.28 \pm 0.03$ and $b = 28.4 \pm 0.5$.  When $\mathcal G \ge 0.8$ the Aboav fit begins to quickly decay, with the average correlation reducing to  $r = 0.35 \pm 0.03$.  The slope parameter $a = 4.9 \pm 0.4$ decreases somewhat but becomes more variable as $\mathcal G$ increases, while the intercept $b = 270 \pm 20$ grows rapidly, due to the high probability of low-sided grains neighboring massive cells.

\begin{figure}
\centering
\includegraphics[width=.5\textwidth]{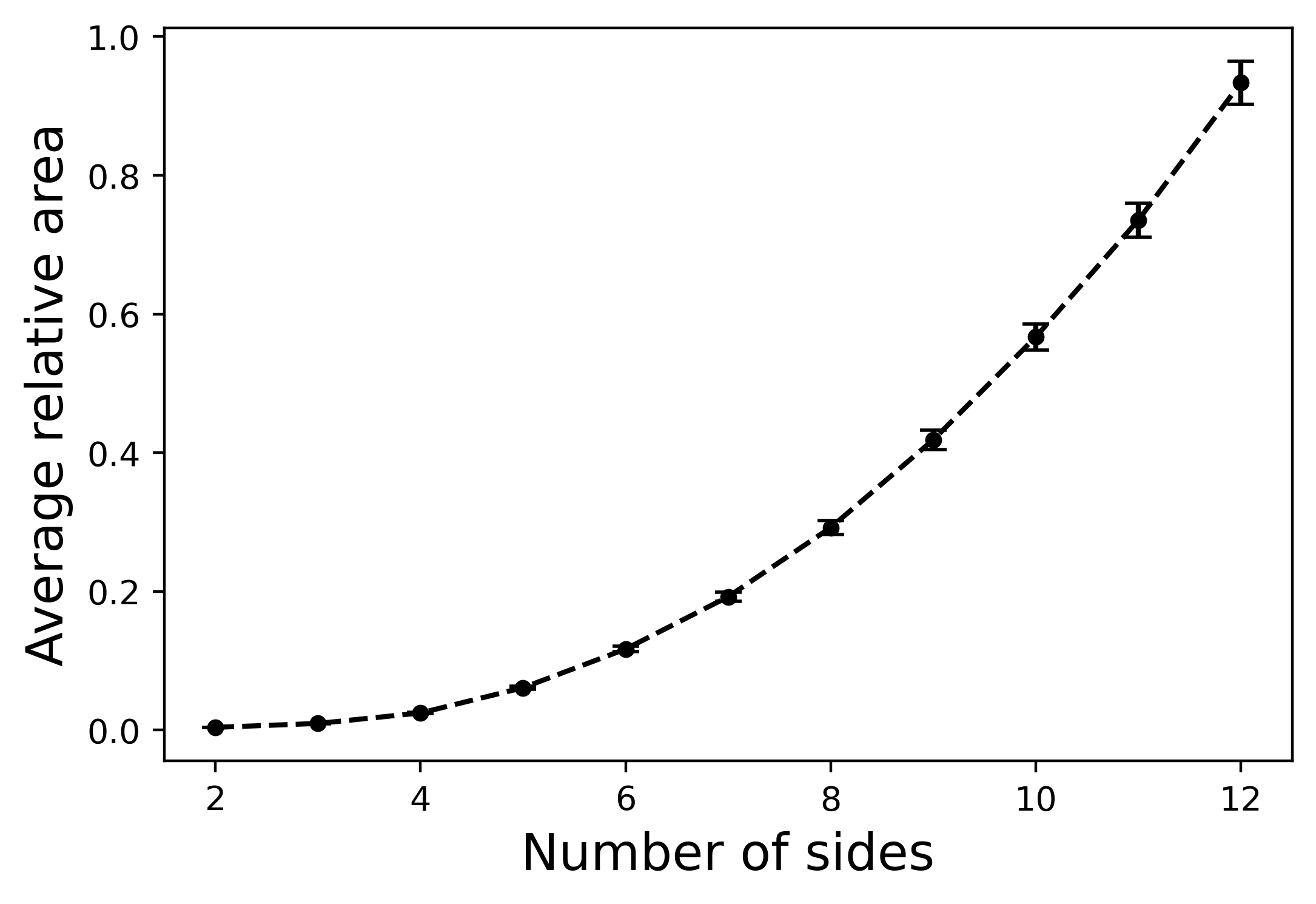}
\caption{Average relative areas $A_n$ of $n$-sided cells for $n = 2,\dots, 12$ over all foam samples, plotted with standard errors.  The quadratic Lewis law fit $A_n =0.011 n^2-0.063 n+ 0.094$ using all data points is shown in the dashed line. } \label{lewisfigall}
\end{figure}

Recall that the Lewis law does not consider area or side correlations for neighboring cells, but rather simply provides average relative areas of $n$-gons.  The generation of massive cells appears to have little effect on the average relative areas on neighboring smaller cells.     This is observed in Fig.\ \ref{lewisfigall}, where we consider relative area data in aggregate for each $n$-sided cell for $n = 2, \dots, 12$.  Variance of relative areas grows between samples as the number of sides increases, with the largest standard error of 0.03 occurring for 12 sided cells.
For each $n = 2, \dots, 12$, we aggregate relative areas across all images, and then compute mean relative areas.  The resulting graph fits well against the quadratic Lewis law $A_n = \beta_2 n^2+ \beta_1 n+ \beta_0$,  with coefficient of determination $R^2 = 0.99$ and fitted values  $\beta_2 = 0.011$, $\beta_1 = -0.063$, and $\beta_0 = 0.094$.

\section{Discussion}  \label{sec:dis}

In this study, we have gathered statistics for a class of two-dimensional rupturing soap foams.  The method for generating and rupturing foams is simple to reproduce.  However, given the stochastic nature of the rupturing process we find variation in area and side number statistics among both heated and unheated foams.  Because we were able to replicate the experiment for 30 different foam samples, we provided error estimates for our statistics.   The foams in this study were restricted to soap foams, but the gelation measures introduced in this work could  be applied to any dynamic two-dimensional planar network.  It would be interesting to investigate if experimental parameters such as viscosity could affect spatial correlations of ruptures and, subsequently, statistics such as the Aboav law.

In Fig.\ \ref{realfoam}(b) and in other foam samples, a Gini coefficient of 0.8 corresponds to when the foam is interspersed with multiple large cells containing tens of neighbors, most of which are convex or approximately convex.  The merging of cells occurs in a continuous manner, so by visual inspection it is difficult to identify a particular Gini value which defines a clear phase transition.  However, Fig.\ \ref{aboavevo} shows that a Gini coefficient of 0.8 serves as an approximate cutoff for when foams no longer follow the linear Aboav law, which is also difficult to verify by inspection alone.     The findings in this paper are empirical, but theoretical models and analytic tools might help explain why such a phenomenon occurs.  In particular, a deeper investigation into gelation behavior might elucidate why a breakdown of the Aboav Law appears when the Gini coefficient becomes large, and whether this breakdown occurs simultaneously with a gelation time in a kinetic model.  In future work, we hope to compare experimental data on side-distributions against the computational rupture model in Ref.\ \cite{klobusicky2021markov} which focused only on topological gelation.  To incorporate areas in kinetic models of foam evolution, we could generalize Eq.\ (\ref{topreact}) to include the merging of cell areas.  However, there exists no known analog of the $n-6$ rule for area evolution in rupturing foams, as the dynamics of comparing cells boundaries before and after rupture is complicated.  One simple approximation may consider an additive model for areas, in which a rupture causes two cells of areas $A_1$ and $A_2$ to merge into a single cell of area $A_1+ A_2$, with all other cells maintaining their original areas. For a foam comprised of cell sides and areas $\{(s_i, a_i)\}_{i = 1}^{N}$, the reaction resulting from a rupture is then 
\begin{align}
(s_i,a_i)+ &(s_j,a_j)+(s_k,a_k)+ (s_l,a_l) \label{areatopreact}\\&\rightharpoonup  (s_i+s_j-4,a_i+a_{j})+ (s_k-1,a_k)+ (s_l-1,a_l).  \nonumber
\end{align}
The limiting integro-differential equation for the evolution of area and side distributions would be unusual in that it would combine a continuous merging for areas with discrete merging for side numbers. A discrete time stochastic particle system following (\ref{areatopreact}) can compare statistics against those found in this study. The Gini coefficient is particularly amenable to such a reduced model, as it only requires a sorted list of cell areas to compute.

Assuming no memory effects, the evolution of cell areas and topologies can be modeled as a homogeneous Markov process.  For a state space consisting of a list of cell sides and areas, transitions would be determined by randomly selecting four cells to undergo the reaction (\ref{areatopreact}).  In most mean field models for grain coarsening,  cells are chosen solely in proportion to their number of sides \cite{marder1987soap,flyvbjerg1993model,fra882} (for a notable exception, Chae and Tabor \cite{chae1997dynamics} chose cells proportional to cell length).   In \cite{klobusicky2020two}, for instance, it was shown that adding first-order topological neighbor correlations in grain coarsening models produced significant differences in statistical topologies.  Incorporating other topological and geometrical relations into transition probabilities may help produce more accurate models of network statistics.   The Aboav laws, which in Sec.\ \ref{sec:fid} were found to hold reasonably well for foams with Gini coefficient less than 0.8, can perhaps be used, although other models will likely be necessary after entering into the non-Aboav, or gelation, regime.

We close by mentioning a class of related coagulation equations which may be amenable to rigorous analysis.  The reaction (\ref{topreact})  can be seen as an instance of a reaction with dissipation, since six sides are lost with each rupture.  In Ref.\ \cite{wattis2004coagulation}, merging reactions for the Smoluchowski equation were paired with an additional annihilation reaction $C_j \rightharpoonup \emptyset$ for clusters size $j \ge 1$, and several exact and approximate formulas were found for distributions of $C_n$.  An analytic study of merging clusters 
\begin{equation}
C_i + C_j \rightharpoonup C_{i + j - \phi(i,j)}, \qquad i,j \ge 1 \nonumber
\end{equation}
 for some \textit{reaction cost} $\phi(i,j)$ remains to be seen. As in Ref.\ \cite{wattis2004coagulation}, a key question is whether the system completely dissipates before producing a gel.

\vspace{10 pt}

\begin{acknowledgements}  The work of J.K.\ and E.O.\ is partially supported by NASA under Grant No.\ 80NSSC20M0097. The work of J.K.\ is partially supported by the National Science Foundation under Grant No.\ 2316289.
\end{acknowledgements}

\bibliography{manuscript}

\end{document}